\documentclass[twocolumn]{aastex631}

\usepackage{xspace}

\newcommand {\nicer}{\textit{NICER}\xspace}
\newcommand {\psrb}{{PSR~B0540-69}\xspace}
\newcommand {\asec}{{$^{\prime \prime}$}\xspace}

\shorttitle{X-ray polarimetry of PSR~B0540--69}
\shortauthors{Xie et al.}

\begin{document}
\title{First detection of polarization in X-rays for PSR~B0540--69 and its nebula}

\correspondingauthor{Fei Xie}
\email{xief@gxu.edu.cn}
\author[0000-0002-0105-5826]{Fei Xie}
\affiliation{Guangxi Key Laboratory for Relativistic Astrophysics, School of Physical Science and Technology, Guangxi University, Nanning 530004, China}
\affiliation{INAF Istituto di Astrofisica e Planetologia Spaziali, Via del Fosso del Cavaliere 100, 00133 Roma, Italy}

\author[0000-0001-6395-2066]{Josephine Wong}
\affiliation{Department of Physics and Kavli Institute for Particle Astrophysics and Cosmology, Stanford University, Stanford, California 94305, USA}
\author[0000-0001-8916-4156]{Fabio La Monaca}
\affiliation{INAF Istituto di Astrofisica e Planetologia Spaziali, Via del Fosso del Cavaliere 100, 00133 Roma, Italy}
\affiliation{Dipartimento di Fisica, Universit\`{a} degli Studi di Roma ``Tor Vergata'', Via della Ricerca Scientifica 1, 00133 Roma, Italy}
\affiliation{Dipartimento di Fisica, Università degli Studi di Roma “La Sapienza”, Piazzale Aldo Moro 5, 00185 Roma, Italy}
\author[0000-0001-6711-3286]{Roger W. Romani}
\affiliation{Department of Physics and Kavli Institute for Particle Astrophysics and Cosmology, Stanford University, Stanford, California 94305, USA}
\author[0000-0001-9739-367X]{Jeremy Heyl}
\affiliation{University of British Columbia, Vancouver, BC V6T 1Z4, Canada}
\author[0000-0002-3638-0637]{Philip Kaaret}
\affiliation{NASA Marshall Space Flight Center, Huntsville, AL 35812, USA}
\author[0000-0003-0331-3259]{Alessandro Di Marco}
\affiliation{INAF Istituto di Astrofisica e Planetologia Spaziali, Via del Fosso del Cavaliere 100, 00133 Roma, Italy}
\author[0000-0002-8848-1392]{Niccol\`{o} Bucciantini}
\affiliation{INAF Osservatorio Astrofisico di Arcetri, Largo Enrico Fermi 5, 50125 Firenze, Italy}
\affiliation{Dipartimento di Fisica e Astronomia, Universit\`{a} degli Studi di Firenze, Via Sansone 1, 50019 Sesto Fiorentino (FI), Italy}
\affiliation{Istituto Nazionale di Fisica Nucleare, Sezione di Firenze, Via Sansone 1, 50019 Sesto Fiorentino (FI), Italy}

\author[0009-0007-8686-9012]{Kuan Liu}
\affiliation{Guangxi Key Laboratory for Relativistic Astrophysics, School of Physical Science and Technology, Guangxi University, Nanning 530004, China}
\author[0000-0002-5847-2612]{Chi-Yung Ng}
\affiliation{Department of Physics, The University of Hong Kong, Pokfulam, Hong Kong}
\author[0000-0002-7574-1298]{Niccol\`{o} Di Lalla}
\affiliation{Department of Physics and Kavli Institute for Particle Astrophysics and Cosmology, Stanford University, Stanford, California 94305, USA}
\author[0000-0002-5270-4240]{Martin C. Weisskopf}
\affiliation{NASA Marshall Space Flight Center, Huntsville, AL 35812, USA}

\author[0000-0003-4925-8523]{Enrico Costa}
\affiliation{INAF Istituto di Astrofisica e Planetologia Spaziali, Via del Fosso del Cavaliere 100, 00133 Roma, Italy}
\author[0000-0002-7781-4104]{Paolo Soffitta}
\affiliation{INAF Istituto di Astrofisica e Planetologia Spaziali, Via del Fosso del Cavaliere 100, 00133 Roma, Italy}
\author[0000-0003-3331-3794]{Fabio Muleri}
\affiliation{INAF Istituto di Astrofisica e Planetologia Spaziali, Via del Fosso del Cavaliere 100, 00133 Roma, Italy}	

\author[0000-0002-4576-9337]{Matteo Bachetti}
\affiliation{INAF Osservatorio Astronomico di Cagliari, Via della Scienza 5, 09047 Selargius (CA), Italy}
\author[0000-0001-7397-8091]{Maura Pilia}
\affiliation{INAF Osservatorio Astronomico di Cagliari, Via della Scienza 5, 09047 Selargius (CA), Italy}
\author[0000-0002-9774-0560]{John Rankin}
\affiliation{INAF Istituto di Astrofisica e Planetologia Spaziali, Via del Fosso del Cavaliere 100, 00133 Roma, Italy}
\author[0000-0003-1533-0283]{Sergio Fabiani}
\affiliation{INAF Istituto di Astrofisica e Planetologia Spaziali, Via del Fosso del Cavaliere 100, 00133 Roma, Italy}

\author[0000-0002-3777-6182]{Iv\'an Agudo}
\affiliation{Instituto de Astrof\'{i}sica de Andaluc\'{i}a -- CSIC, Glorieta de la Astronom\'{i}a s/n, 18008 Granada, Spain}
\author[0000-0002-5037-9034]{Lucio A. Antonelli}
\affiliation{INAF Osservatorio Astronomico di Roma, Via Frascati 33, 00040 Monte Porzio Catone (RM), Italy}
\affiliation{Space Science Data Center, Agenzia Spaziale Italiana, Via del Politecnico snc, 00133 Roma, Italy}
\author[0000-0002-9785-7726]{Luca Baldini}
\affiliation{Istituto Nazionale di Fisica Nucleare, Sezione di Pisa, Largo B. Pontecorvo 3, 56127 Pisa, Italy}
\affiliation{Dipartimento di Fisica, Universit\`{a} di Pisa, Largo B. Pontecorvo 3, 56127 Pisa, Italy}
\author[0000-0002-5106-0463]{Wayne H. Baumgartner}
\affiliation{NASA Marshall Space Flight Center, Huntsville, AL 35812, USA}
\author[0000-0002-2469-7063]{Ronaldo Bellazzini}
\affiliation{Istituto Nazionale di Fisica Nucleare, Sezione di Pisa, Largo B. Pontecorvo 3, 56127 Pisa, Italy}
\author[0000-0002-4622-4240]{Stefano Bianchi}
\affiliation{Dipartimento di Matematica e Fisica, Universit\`{a} degli Studi Roma Tre, Via della Vasca Navale 84, 00146 Roma, Italy}
\author[0000-0002-0901-2097]{Stephen D. Bongiorno}
\affiliation{NASA Marshall Space Flight Center, Huntsville, AL 35812, USA}
\author[0000-0002-4264-1215]{Raffaella Bonino}
\affiliation{Istituto Nazionale di Fisica Nucleare, Sezione di Torino, Via Pietro Giuria 1, 10125 Torino, Italy}
\affiliation{Dipartimento di Fisica, Universit\`{a} degli Studi di Torino, Via Pietro Giuria 1, 10125 Torino, Italy}
\author[0000-0002-9460-1821]{Alessandro Brez}
\affiliation{Istituto Nazionale di Fisica Nucleare, Sezione di Pisa, Largo B. Pontecorvo 3, 56127 Pisa, Italy}
\author[0000-0002-6384-3027]{Fiamma Capitanio}
\affiliation{INAF Istituto di Astrofisica e Planetologia Spaziali, Via del Fosso del Cavaliere 100, 00133 Roma, Italy}
\author[0000-0003-1111-4292]{Simone Castellano}
\affiliation{Istituto Nazionale di Fisica Nucleare, Sezione di Pisa, Largo B. Pontecorvo 3, 56127 Pisa, Italy}
\author[0000-0001-7150-9638]{Elisabetta Cavazzuti}
\affiliation{Agenzia Spaziale Italiana, Via del Politecnico snc, 00133 Roma, Italy}
\author[0000-0002-4945-5079]{Chien-Ting Chen}
\affiliation{Science and Technology Institute, Universities Space Research Association, Huntsville, AL 35805, USA}
\author[0000-0002-0712-2479]{Stefano Ciprini}
\affiliation{Istituto Nazionale di Fisica Nucleare, Sezione di Roma ``Tor Vergata'', Via della Ricerca Scientifica 1, 00133 Roma, Italy}
\affiliation{Space Science Data Center, Agenzia Spaziale Italiana, Via del Politecnico snc, 00133 Roma, Italy}
\author[0000-0001-5668-6863]{Alessandra De Rosa}
\affiliation{INAF Istituto di Astrofisica e Planetologia Spaziali, Via del Fosso del Cavaliere 100, 00133 Roma, Italy}
\author[0000-0002-3013-6334]{Ettore Del Monte}
\affiliation{INAF Istituto di Astrofisica e Planetologia Spaziali, Via del Fosso del Cavaliere 100, 00133 Roma, Italy}
\author[0000-0002-5614-5028]{Laura Di Gesu}
\affiliation{Agenzia Spaziale Italiana, Via del Politecnico snc, 00133 Roma, Italy}
\author[0000-0002-4700-4549]{Immacolata Donnarumma}
\affiliation{Agenzia Spaziale Italiana, Via del Politecnico snc, 00133 Roma, Italy}
\author[0000-0001-8162-1105]{Victor Doroshenko}
\affiliation{Institut f\"{u}r Astronomie und Astrophysik, Universität Tübingen, Sand 1, 72076 T\"{u}bingen, Germany}
\author[0000-0003-0079-1239]{Michal Dov\v{c}iak}
\affiliation{Astronomical Institute of the Czech Academy of Sciences, Bo\v{c}n\'{i} II 1401/1, 14100 Praha 4, Czech Republic}
\author[0000-0003-4420-2838]{Steven R. Ehlert}
\affiliation{NASA Marshall Space Flight Center, Huntsville, AL 35812, USA}
\author[0000-0003-1244-3100]{Teruaki Enoto}
\affiliation{RIKEN Cluster for Pioneering Research, 2-1 Hirosawa, Wako, Saitama 351-0198, Japan}
\author[0000-0001-6096-6710]{Yuri Evangelista}
\affiliation{INAF Istituto di Astrofisica e Planetologia Spaziali, Via del Fosso del Cavaliere 100, 00133 Roma, Italy}
\author[0000-0003-1074-8605]{Riccardo Ferrazzoli}
\affiliation{INAF Istituto di Astrofisica e Planetologia Spaziali, Via del Fosso del Cavaliere 100, 00133 Roma, Italy}
\author[0000-0003-3828-2448]{Javier A. Garcia}
\affiliation{NASA Goddard Space Flight Center, Greenbelt, MD 20771, USA}
\author[0000-0002-5881-2445]{Shuichi Gunji}
\affiliation{Yamagata University,1-4-12 Kojirakawa-machi, Yamagata-shi 990-8560, Japan}
\author{Kiyoshi Hayashida}
\altaffiliation{Deceased}
\affiliation{Osaka University, 1-1 Yamadaoka, Suita, Osaka 565-0871, Japan}
\author[0000-0002-0207-9010]{Wataru Iwakiri}
\affiliation{International Center for Hadron Astrophysics, Chiba University, Chiba 263-8522, Japan}
\author[0000-0001-9522-5453]{Svetlana G. Jorstad}
\affiliation{Institute for Astrophysical Research, Boston University, 725 Commonwealth Avenue, Boston, MA 02215, USA}
\affiliation{Department of Astrophysics, St. Petersburg State University, Universitetsky pr. 28, Petrodvoretz, 198504 St. Petersburg, Russia}
\author[0000-0002-5760-0459]{Vladimir Karas}
\affiliation{Astronomical Institute of the Czech Academy of Sciences, Bo\v{c}n\'{i} II 1401/1, 14100 Praha 4, Czech Republic}
\author[0000-0001-7477-0380]{Fabian Kislat}
\affiliation{Department of Physics and Astronomy and Space Science Center, University of New Hampshire, Durham, NH 03824, USA}
\author{Takao Kitaguchi}
\affiliation{RIKEN Cluster for Pioneering Research, 2-1 Hirosawa, Wako, Saitama 351-0198, Japan}
\author[0000-0002-0110-6136]{Jeffery J. Kolodziejczak}
\affiliation{NASA Marshall Space Flight Center, Huntsville, AL 35812, USA}
\author[0000-0002-1084-6507]{Henric Krawczynski}
\affiliation{Physics Department and McDonnell Center for the Space Sciences, Washington University in St. Louis, St. Louis, MO 63130, USA}
\author[0000-0002-0984-1856]{Luca Latronico}
\affiliation{Istituto Nazionale di Fisica Nucleare, Sezione di Torino, Via Pietro Giuria 1, 10125 Torino, Italy}
\author[0000-0001-9200-4006]{Ioannis Liodakis}
\affiliation{NASA Marshall Space Flight Center, Huntsville, AL 35812, USA}
\author[0000-0002-0698-4421]{Simone Maldera}
\affiliation{Istituto Nazionale di Fisica Nucleare, Sezione di Torino, Via Pietro Giuria 1, 10125 Torino, Italy}
\author[0000-0002-0998-4953]{Alberto Manfreda}  
\affiliation{Istituto Nazionale di Fisica Nucleare, Sezione di Napoli, Strada Comunale Cinthia, 80126 Napoli, Italy}
\author[0000-0003-4952-0835]{Fr\'{e}d\'{e}ric Marin}
\affiliation{Universit\'{e} de Strasbourg, CNRS, Observatoire Astronomique de Strasbourg, UMR 7550, 67000 Strasbourg, France}
\author[0000-0002-2055-4946]{Andrea Marinucci}
\affiliation{Agenzia Spaziale Italiana, Via del Politecnico snc, 00133 Roma, Italy}
\author[0000-0001-7396-3332]{Alan P. Marscher}
\affiliation{Institute for Astrophysical Research, Boston University, 725 Commonwealth Avenue, Boston, MA 02215, USA}
\author[0000-0002-6492-1293]{Herman L. Marshall}
\affiliation{MIT Kavli Institute for Astrophysics and Space Research, Massachusetts Institute of Technology, 77 Massachusetts Avenue, Cambridge, MA 02139, USA}
\author[0000-0002-1704-9850]{Francesco Massaro}
\affiliation{Istituto Nazionale di Fisica Nucleare, Sezione di Torino, Via Pietro Giuria 1, 10125 Torino, Italy}
\affiliation{Dipartimento di Fisica, Universit\`{a} degli Studi di Torino, Via Pietro Giuria 1, 10125 Torino, Italy}
\author[0000-0002-2152-0916]{Giorgio Matt}
\affiliation{Dipartimento di Matematica e Fisica, Universit\`{a} degli Studi Roma Tre, Via della Vasca Navale 84, 00146 Roma, Italy}
\author{Ikuyuki Mitsuishi}
\affiliation{Graduate School of Science, Division of Particle and Astrophysical Science, Nagoya University, Furo-cho, Chikusa-ku, Nagoya, Aichi 464-8602, Japan}
\author[0000-0001-7263-0296]{Tsunefumi Mizuno}
\affiliation{Hiroshima Astrophysical Science Center, Hiroshima University, 1-3-1 Kagamiyama, Higashi-Hiroshima, Hiroshima 739-8526, Japan}
\author[0000-0002-6548-5622]{Michela Negro} 
\affiliation{Department of Physics and Astronomy, Louisiana State University, Baton Rouge, LA 70803, USA}
\author[0000-0002-1868-8056]{Stephen L. O'Dell}
\affiliation{NASA Marshall Space Flight Center, Huntsville, AL 35812, USA}
\author[0000-0002-5448-7577]{Nicola Omodei}
\affiliation{Department of Physics and Kavli Institute for Particle Astrophysics and Cosmology, Stanford University, Stanford, California 94305, USA}
\author[0000-0001-6194-4601]{Chiara Oppedisano}
\affiliation{Istituto Nazionale di Fisica Nucleare, Sezione di Torino, Via Pietro Giuria 1, 10125 Torino, Italy}
\author[0000-0001-6289-7413]{Alessandro Papitto}
\affiliation{INAF Osservatorio Astronomico di Roma, Via Frascati 33, 00040 Monte Porzio Catone (RM), Italy}
\author[0000-0002-7481-5259]{George G. Pavlov}
\affiliation{Department of Astronomy and Astrophysics, Pennsylvania State University, University Park, PA 16802, USA}
\author[0000-0001-6292-1911]{Abel L. Peirson}
\affiliation{Department of Physics and Kavli Institute for Particle Astrophysics and Cosmology, Stanford University, Stanford, California 94305, USA}
\author[0000-0003-3613-4409]{Matteo Perri}
\affiliation{Space Science Data Center, Agenzia Spaziale Italiana, Via del Politecnico snc, 00133 Roma, Italy}
\affiliation{INAF Osservatorio Astronomico di Roma, Via Frascati 33, 00040 Monte Porzio Catone (RM), Italy}
\author[0000-0003-1790-8018]{Melissa Pesce-Rollins}
\affiliation{Istituto Nazionale di Fisica Nucleare, Sezione di Pisa, Largo B. Pontecorvo 3, 56127 Pisa, Italy}
\author[0000-0001-6061-3480]{Pierre-Olivier Petrucci}
\affiliation{Universit\'{e} Grenoble Alpes, CNRS, IPAG, 38000 Grenoble, France}
\author[0000-0001-5902-3731]{Andrea Possenti}
\affiliation{INAF Osservatorio Astronomico di Cagliari, Via della Scienza 5, 09047 Selargius (CA), Italy}
\author[0000-0002-0983-0049]{Juri Poutanen}
\affiliation{Department of Physics and Astronomy,  20014 University of Turku, Finland}
\author[0000-0002-2734-7835]{Simonetta Puccetti}
\affiliation{Space Science Data Center, Agenzia Spaziale Italiana, Via del Politecnico snc, 00133 Roma, Italy}
\author[0000-0003-1548-1524]{Brian D. Ramsey}
\affiliation{NASA Marshall Space Flight Center, Huntsville, AL 35812, USA}
\author[0000-0003-0411-4243]{Ajay Ratheesh}
\affiliation{INAF Istituto di Astrofisica e Planetologia Spaziali, Via del Fosso del Cavaliere 100, 00133 Roma, Italy}
\author[0000-0002-7150-9061]{Oliver J. Roberts}
\affiliation{Science and Technology Institute, Universities Space Research Association, Huntsville, AL 35805, USA}
\author[0000-0001-5676-6214]{Carmelo Sgr\`{o}}
\affiliation{Istituto Nazionale di Fisica Nucleare, Sezione di Pisa, Largo B. Pontecorvo 3, 56127 Pisa, Italy}
\author[0000-0002-6986-6756]{Patrick Slane}
\affiliation{Center for Astrophysics, Harvard \& Smithsonian, 60 Garden St, Cambridge, MA 02138, USA}
\author[0000-0003-0802-3453]{Gloria Spandre}
\affiliation{Istituto Nazionale di Fisica Nucleare, Sezione di Pisa, Largo B. Pontecorvo 3, 56127 Pisa, Italy}
\author[0000-0002-2954-4461]{Douglas A. Swartz}
\affiliation{Science and Technology Institute, Universities Space Research Association, Huntsville, AL 35805, USA}
\author[0000-0002-8801-6263]{Toru Tamagawa}
\affiliation{RIKEN Cluster for Pioneering Research, 2-1 Hirosawa, Wako, Saitama 351-0198, Japan}
\author[0000-0003-0256-0995]{Fabrizio Tavecchio}
\affiliation{INAF Osservatorio Astronomico di Brera, via E. Bianchi 46, 23807 Merate (LC), Italy}
\author[0000-0002-1768-618X]{Roberto Taverna}
\affiliation{Dipartimento di Fisica e Astronomia, Universit\`{a} degli Studi di Padova, Via Marzolo 8, 35131 Padova, Italy}
\author{Yuzuru Tawara}
\affiliation{Graduate School of Science, Division of Particle and Astrophysical Science, Nagoya University, Furo-cho, Chikusa-ku, Nagoya, Aichi 464-8602, Japan}
\author[0000-0002-9443-6774]{Allyn F. Tennant}
\affiliation{NASA Marshall Space Flight Center, Huntsville, AL 35812, USA}
\author[0000-0003-0411-4606]{Nicholas E. Thomas}
\affiliation{NASA Marshall Space Flight Center, Huntsville, AL 35812, USA}
\author[0000-0002-6562-8654]{Francesco Tombesi}
\affiliation{Dipartimento di Fisica, Universit\`{a} degli Studi di Roma ``Tor Vergata'', Via della Ricerca Scientifica 1, 00133 Roma, Italy}
\affiliation{Istituto Nazionale di Fisica Nucleare, Sezione di Roma ``Tor Vergata'', Via della Ricerca Scientifica 1, 00133 Roma, Italy}
\affiliation{Department of Astronomy, University of Maryland, College Park, Maryland 20742, USA}
\author[0000-0002-3180-6002]{Alessio Trois}
\affiliation{INAF Osservatorio Astronomico di Cagliari, Via della Scienza 5, 09047 Selargius (CA), Italy}
\author[0000-0002-9679-0793]{Sergey S. Tsygankov}
\affiliation{Department of Physics and Astronomy,  20014 University of Turku, Finland}
\author[0000-0003-3977-8760]{Roberto Turolla}
\affiliation{Dipartimento di Fisica e Astronomia, Universit\`{a} degli Studi di Padova, Via Marzolo 8, 35131 Padova, Italy}
\affiliation{Mullard Space Science Laboratory, University College London, Holmbury St Mary, Dorking, Surrey RH5 6NT, UK}
\author[0000-0002-4708-4219]{Jacco Vink}
\affiliation{Anton Pannekoek Institute for Astronomy \& GRAPPA, University of Amsterdam, Science Park 904, 1098 XH Amsterdam, The Netherlands}
\author[0000-0002-7568-8765]{Kinwah Wu}
\affiliation{Mullard Space Science Laboratory, University College London, Holmbury St Mary, Dorking, Surrey RH5 6NT, UK}
\author[0000-0001-5326-880X]{Silvia Zane}
\affiliation{Mullard Space Science Laboratory, University College London, Holmbury St Mary, Dorking, Surrey RH5 6NT, UK}

\collaboration{98}{(IXPE Collaboration)}

\author[0000-0002-9249-0515]{Zorawar Wadiasingh}
\affiliation{Astrophysics Science Division, NASA Goddard Space Flight Center, 8800 Greenbelt Road, Greenbelt, MD, 20771, USA}
\affiliation{Department of Astronomy, University of Maryland College Park, 4296 Stadium Dr., PSC, College Park, MD, 20742, USA}
\affiliation{Center for Research and Exploration in Space Science and Technology, NASA/GSFC, 8800 Greenbelt Road, Greenbelt, MD, 20771, USA}
\author[0000-0002-6089-6836]{Wynn C. G. Ho}
\affiliation{Department of Physics and Astronomy, Haverford College, 370 Lancaster Avenue, Haverford, PA, 19041, USA}
\author[0000-0001-6119-859X]{Alice K. Harding}
\affiliation{Theoretical Division, Los Alamos National Laboratory, Los Alamos, NM 87545, USA}
\author[0000-0001-7115-2819]{Keith C. Gendreau}
\affiliation{Astrophysics Science Division, NASA Goddard Space Flight Center, 8800 Greenbelt Road, Greenbelt, MD, 20771, USA}
\author{Zaven Arzoumanian}
\affiliation{Astrophysics Science Division, NASA Goddard Space Flight Center, 8800 Greenbelt Road, Greenbelt, MD, 20771, USA}



\begin{abstract}
We report on X-ray polarization measurements of the extra-galactic Crab-like \psrb and its Pulsar Wind Nebula (PWN) in the Large Magellanic Cloud (LMC), using a $\sim$ 850 ks Imaging X-ray Polarimetry Explorer (IXPE) exposure. The PWN is unresolved by IXPE. No statistically significant polarization is detected for the image-averaged data, giving a 99\% confidence polarization upper limit (MDP$_{99}$) of 5.3\% in 2--8 keV energy range. However, a phase-resolved analysis detects polarization for both the nebula and pulsar in the 4--6 keV energy range. For the PWN defined as the off-pulse phases, the polarization degree (PD) of $(24.5\pm 5.3)\%$ and polarization angle (PA) of $(78.1\pm6.2)^{\circ}$ is detected at 4.6$\sigma$ significance level, consistent with the PA observed in the optical band. In a single on-pulse window, a hint of polarization is measured at 3.8$\sigma$ with polarization degree of $(50.0\pm13.1)\%$ and polarization angle of $(6.2\pm7.4)^\circ$. A `simultaneous' PSR/PWN analysis finds two bins at the edges of the pulse exceeding 3$\sigma$ PD significance, with PD of $(68\pm20)\%$ and $(62\pm20)\%$; intervening bins at 2--3$\sigma$ significance have lower PD, hinting at additional polarization structure.

\end{abstract}
\keywords{pulsar wind nebula, pulsar, polarization -- pulsars: individual (PSR B0540-69)}

\section{Introduction} \label{sec:intro}
\psrb (also known as PSR J0540-6919) is a young Crab-like pulsar located inside the supernova remnant SNR B0540-69.3 in the Large Magellanic Cloud (LMC) satellite galaxy of the Milky Way at a distance of $\sim$ 50\,kpc. It was discovered in the early 1980s by the \textit{Einstein} X-ray Observatory \citep{1984ApJ...287L..19S}. It is the first extragalactic pulsar observed to emit giant radio pulses \citep{2003ApJ...590L..95J} and the first gamma-ray pulsar detected in another galaxy \citep{2015Sci...350..801F}.

\psrb has a short rotation period of 50\,ms, a characteristic age of $\sim$1500\,yr, and a rotational energy loss of ${\dot E} \sim$10$^{38}$ erg s$^{-1}$.  The X-ray pulse profile is double-peaked and asymmetric, with a component separation of $\sim$0.2 in phase \citep{dePlaa2003pulse}, consistent with that in the optical band. Like the Crab, \psrb is also embedded in a bright pulsar wind nebula (PWN) visible at wavelengths from the radio to the X-rays. The optical nebula has a half-power diameter of $\sim$4\asec \citep{1984ApJ...287L..23C}, and the X-ray nebula has an angular diameter of $2^{\prime \prime}-3^{\prime \prime}$ \citep{2001ApJ...546.1159K}. The PWN morphology resembles the Crab, having a torus and jets \citep{2000ApJ...532L.117G}, and is extended along a northeast-southwest axis. The overall X-ray spectrum of \psrb and its nebula is well characterized by a power law with a photon index of $1.92\pm0.11$ \citep{2001ApJ...546.1159K}, as expected if the emission is predominately non-thermal. The pulsar has an index of $1.83\pm0.13$, harder than that of the nebula only $2.09\pm0.14$ \citep{2001ApJ...546.1159K}.

Limited polarization results have been reported for \psrb and its nebula. For the pulsar, we have only optical polarization values. 
\cite{2010A&A...515A.110M} report a pulsar phase-averaged polarization PD=$(16\pm 4)\%$ with an orientation of PA=$(22\pm12)^\circ$, consistent with the semi-major axis of the PWN. While \cite{2011MNRAS.413..611L} report that the pulsar itself had a lower polarization of $(5\pm 2)\%$, and the difference of the pulsar PD values could originate from nebular contamination. The source is faint enough that phase-resolved optical polarimetry has not been obtained.
For the nebula, \cite{1990ApJ...352..167C} report the linear polarization in the optical (V band) integrated over the nebula (within $5.4^{\prime \prime}\times5.4^{\prime \prime}$) of PD=$(5.6\pm1.0)\%$, oriented at an angle of $(79\pm5)^\circ$ east of north. 
In the radio band, \cite{2002ASPC..271..195D} reported a PD of 20\% at 3.5\,cm, 8\% at 6\,cm, and 4.5\% at 20\,cm with position angle of about 80$^\circ$, consistent with the \cite{1990ApJ...352..167C} optical PA.

\psrb is the fourth PWN observed by IXPE, after the Crab \citep{2023NatAs...7..602B}, Vela \citep{2022Natur.612..658X}, and MSH 15-52 \citep{2023arXiv230916067R} and it is the first extra-galactic example. With the PWN unresolved by the 30$^{\prime \prime}$ half-power diameter (HPD) IXPE resolution \citep{Weisskopf2022}, careful phase-resolved analysis is important for \psrb. Here, we report on the first measurements of X-ray polarization from \psrb, with significant detections for both PSR and PWN.

\section{X-ray observations and data reduction}
\subsection{IXPE data}
The Imaging X-ray Polarimetry Explorer (IXPE) is a NASA mission in partnership with the Italian Space Agency launched on 2021 December 9 \citep{Weisskopf2022}. The spacecraft hosts three identical grazing incidence telescopes, providing imaging, timing, and spectral polarimetry in the 2–8 keV nominal energy band. Each telescope has a polarization-sensitive detector unit (DU) equipped with a gas-pixel detector (GPD) \citep{Costa2001, Soffitta2021} placed in the focal plane of an X-ray mirror assembly module (MMA). \psrb has been observed by IXPE in three different periods: (1) December 29 2022 to January 5 2023, (2) January 21--27 2023, and (3) May 10--12 2023, for a total exposure of $\sim$ 850 ks. \psrb observations were released into two data sets at the HEASARC, with the first two observations integrated into OBSID 02001299 and the observation in May as OBSID 02008801. 

Data were extracted and analyzed with the IXPE-dedicated software \textsc{ixpeobssim} \citep{Baldini2022} (v.30.3.0) and HEASOFT 6.31.1 using the Calibration database released on November 17, 2022. Data cuts were used to reduce background events, following the procedure reported in \cite{Di_Marco2023}, and we filtered the good time intervals (GTI) to reduce particle events due to solar activity. This removed 2--3\% of the events in each of the three DUs. For faint sources, such as \psrb, the remaining background is still a substantial fraction of the source flux, especially at high energy, and must still be subtracted in the analysis \citep{Di_Marco2023}, as detailed below.

\subsection{NICER data}
In December 2022, IXPE experienced a timing anomaly. This affected only the first observation of \psrb; timing was restored to normal after a restart. To help define a high accuracy ephemeris for phase-resolved analysis, we also have simultaneous Neutron star Interior Composition Explorer (\nicer) observations. Observations using \nicer were made between January 19, 2023 at 20:46:20 and May 3, 2023 at 19:35:50. These observations spanned the 2023 IXPE observation interval. To enhance the precision of the ephemeris, we included long-term \nicer data observed back to April 2019. For our \nicer analysis, we utilized Level 2 data retrieved from the HEASARC data archive. The total exposure time for the cleaned event file from 7 MPU detectors amounted to 27.5 ks.

\section{Timing analysis of \psrb}
Timing analysis of \psrb was performed including \nicer data, relying on \nicer's exceptional timing accuracy and good coverage during 2023. Barycentric corrections for both the IXPE and \nicer events were made using the \texttt{barycorr} tool in HEASoft v6.31.1. The JPL-DE430 solar-system ephemeris was utilized, with the position of the source set at $\mathrm{R.A.}=05^{\mathrm{h}}40^{\mathrm{m}}10.84^{\mathrm{s}}$ and $\mathrm{Decl.}=-69^{\circ}19'54.2''$ (J2000) according to SIMBAD Astronomical Database \footnote{https://simbad.u-strasbg.fr/simbad/}. To identify the pulsar signal, we employed the $Z^2$ statistic search implemented in \texttt{Stingray} \citep{huppenkothen2019stingray}. For each individual observation, we selected the period that produced the most significant folded pulse profile. 
To obtain a timing solution for \psrb's 2023 observations, we employed a phase-coherent timing analysis. The time of arrivals (ToAs) for the observed pulse profiles were determined by measuring the peak phase in the folded profile. This was accomplished by cross-correlating each profile with the standard profile obtained during \nicer's long-term monitoring of \psrb. The same procedures were applied to the IXPE data. To obtain the timing solution, we used the \texttt{TEMPO2} software \citep{hobbs2006tempo2}, fitting both the \nicer and IXPE TOAs.

To compensate for the 2022 IXPE timing anomaly, we incorporated a time delay parameter (referred to as `JUMP' in TEMPO2) as a freely fitted TOA offset for the first IXPE observation. The results of the best fit are presented in Table \ref{tab:ephemeris} as Ephemeris 1. In addition to the timing anomaly jump, we see time delays of a few milliseconds compared to \nicer in the second and third IXPE observations. These were modeled with additional `JUMP' parameters, resulting in the timing solution shown in Table \ref{tab:ephemeris} as Ephemeris 2.

\begin{figure}[htbp]
\centering
\includegraphics[width=\linewidth]{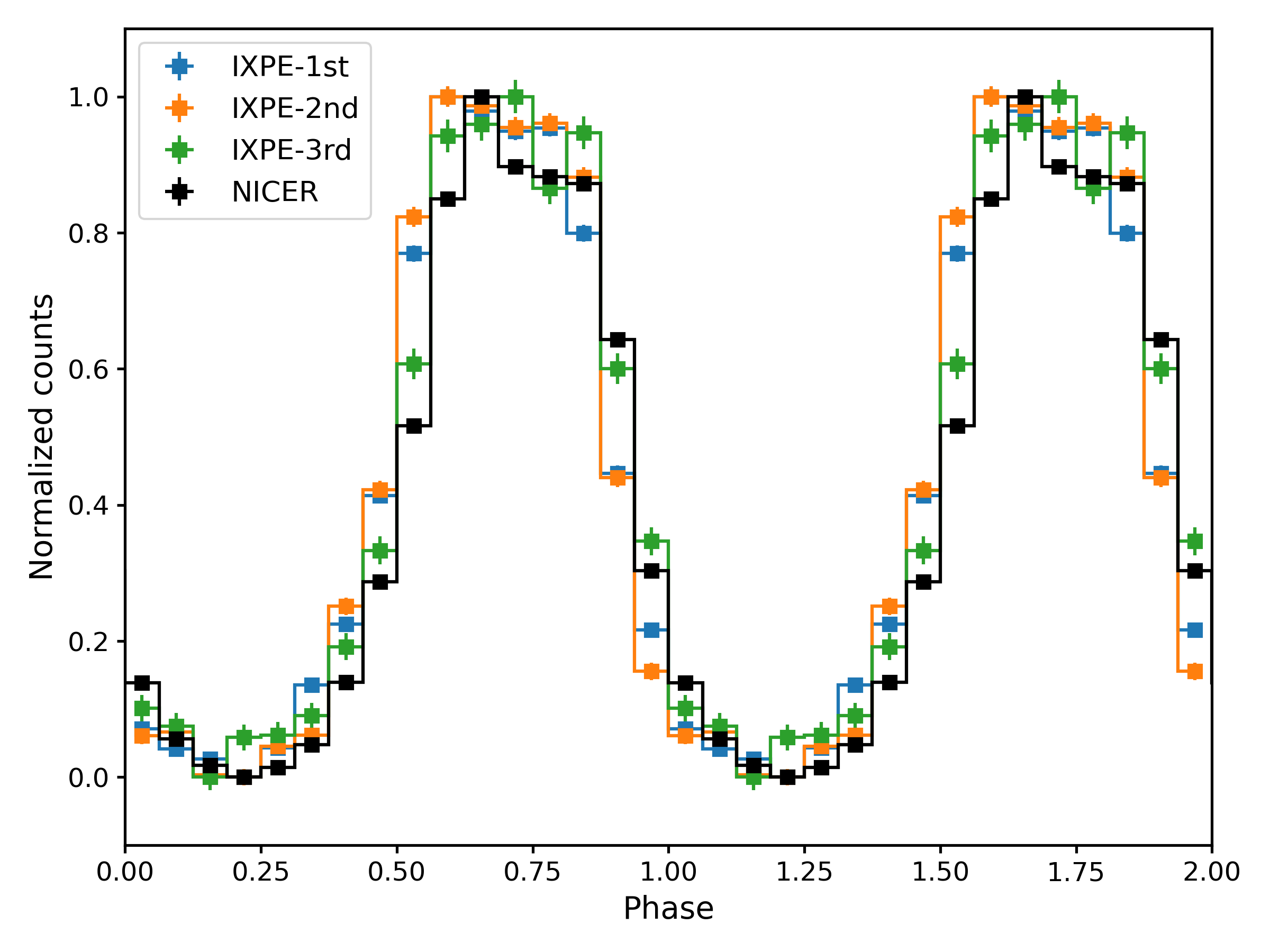}
\caption{Profile of \psrb obtained from the three IXPE observations (in blue, orange, and green respectively) and \nicer summed-up observations (in black) using Ephemeris 2 in Table~\ref{tab:ephemeris}.}
\label{fig:ixpe_nicer_profile}
\end{figure}

We used Ephemeris 2 in our phase-resolved data analysis. The resulting combined \nicer profile and profiles from the three IXPE observations are shown in Fig.~\ref{fig:ixpe_nicer_profile}.

\begin{table}[ht]
\centering
\caption{\label{tab:ephemeris}The ephemeris of \psrb obtained by \nicer and IXPE observations. The Ephemeris 1 is the best-fit timing resolution, where only a single time delay was introduced for the first observation of IXPE. In the Ephemeris 2, time delays were incorporated for all observations of IXPE.}
\begin{tabular}{lcc}
\hline
Parameters & Ephemeris 1 & Ephemeris 2 \\
\hline
PEPOCH (MJD) & 58920 (fixed) & 58920 (fixed) \\
$\nu \,(\mathrm{Hz})$  & 19.660547(3)  & 19.660545(3)  \\
$\dot{\nu}\,(10^{-10}\,\mathrm{Hz}\cdot\mathrm{s}^{-1})$ & -2.5287(6) & -2.5281(6) \\
$\ddot{\nu}\,(10^{-21}\,\mathrm{Hz}\cdot\mathrm{s}^{-2})$ & 7.1(7) & 6.5(6) \\
JUMP1 (s)\footnote{the three JUMP parameters are the time delay of three IXPE observations relative to \nicer} & -0.029(1) & -0.0283(8) \\
JUMP2 (s) & - & 0.0027(3) \\
JUMP3 (s) & - & 0.0024(8) \\
\hline
\end{tabular}
\end{table}

\begin{figure}[htbp]
\centering
\includegraphics[width=\linewidth]{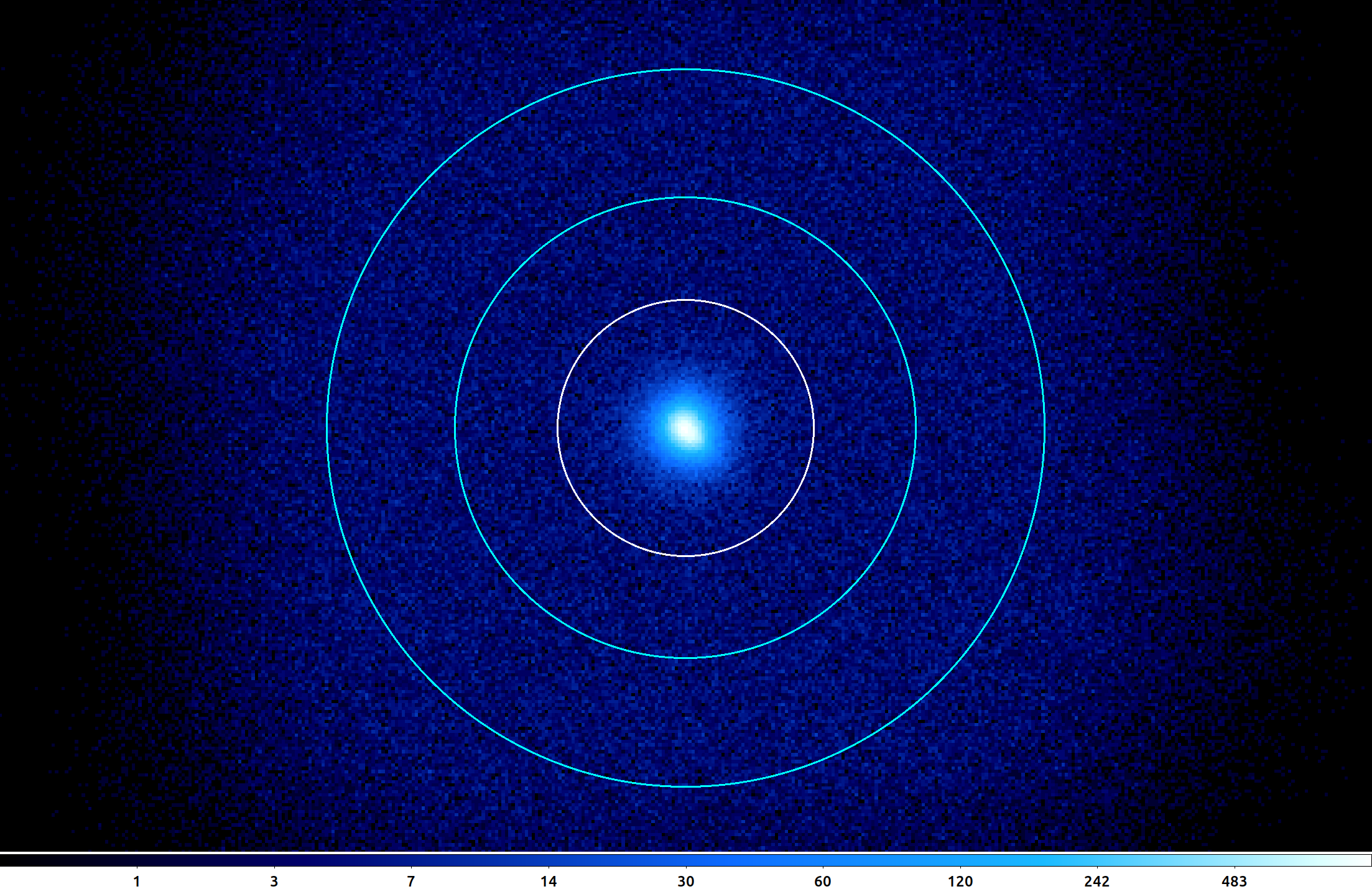}
\caption{Total nebula source (white 100$^{\prime \prime}$ radius circle) and background (cyan annulus, inner radius 180$^{\prime \prime}$, outer radius 280$^{\prime \prime}$) regions shown on images from DU1. Intensity is on a logarithmic scale to bring out the faint background.}
\label{fig:bkg_region}
\end{figure}

\section{Polarimetric analysis}

\begin{figure*}
\centering
\includegraphics[width=0.75\textwidth]{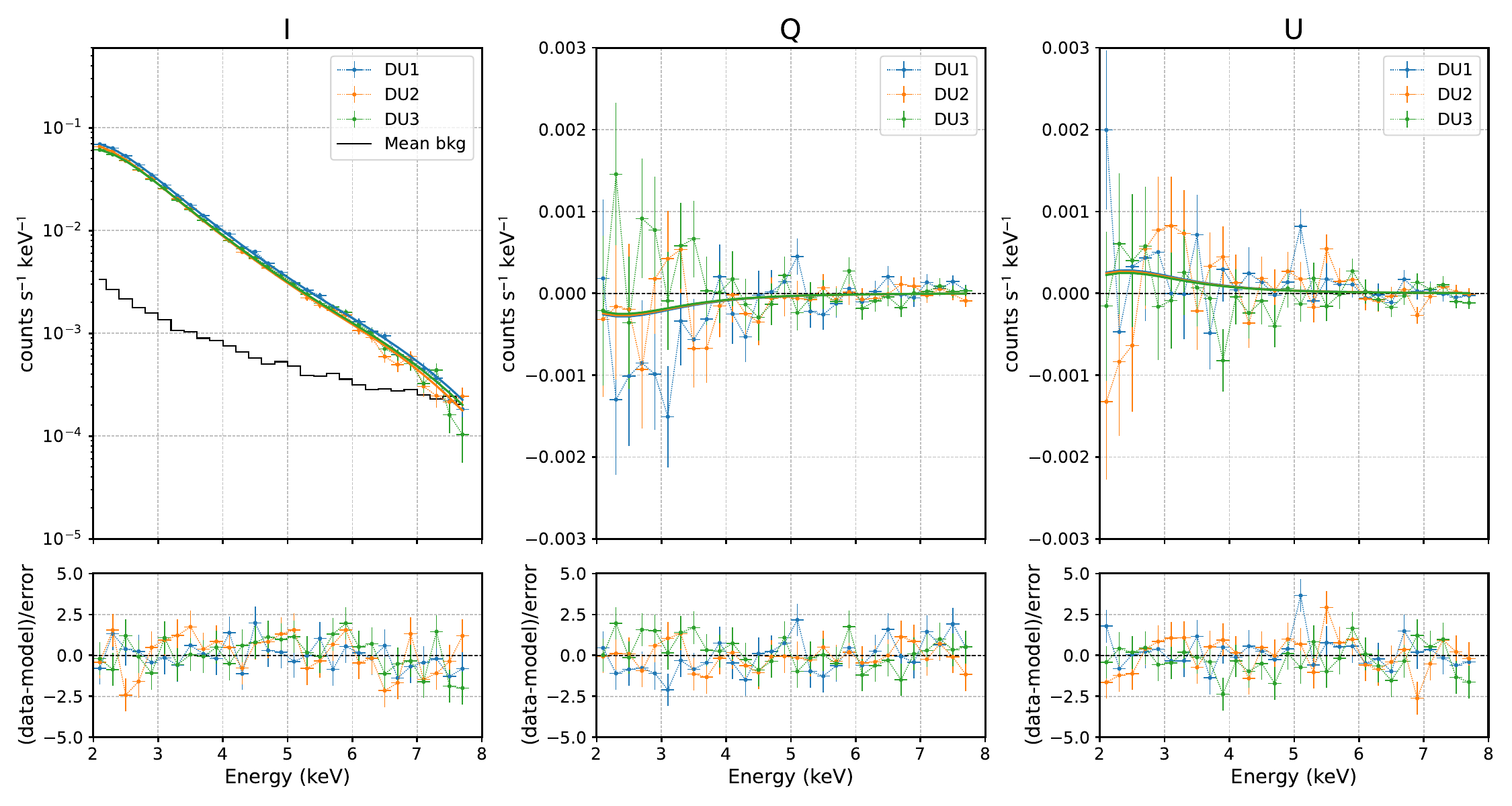}
\caption{Spectral joint fit for the phase-averaged $I$, $Q$, and $U$ Stokes fluxes in the 2--8 keV energy band using three IXPE detectors and the \texttt{const*tbabs*polpow} model. Fit values are tabulated in Table~\ref{tab:xspec_par}. The average background $I$ spectrum is reported in black.}
\label{fig:spectra_q_u}
\end{figure*}

\subsection{Phase-averaged analysis}
As noted, the $3^{\prime \prime}$ angular extent of the \psrb X-ray nebula is much smaller than the 30$^{\prime \prime}$ IXPE resolution. The surrounding X-ray SNR extends to 60$^{\prime \prime}$, but is fainter and softer and contributes only weakly to the IXPE flux. We therefore start by analyzing the polarized properties of the integrated \psrb complex.

The source photons were extracted from a circular region with a radius of 100$^{\prime \prime}$, and the background photons from an annular region with inner and outer radii of 180$^{\prime \prime}$ and 280$^{\prime \prime}$ respectively, both centered on the pulsar, as shown in Fig.~\ref{fig:bkg_region}. The background region was chosen to avoid the edge of the instrument field of view \citep[see][]{Di_Marco2023}. This analysis is performed both with the model-independent PCUBE algorithm in \textsc{ixpeobssim} software \citep{Baldini2022} and with \textsc{xspec} (v.12.13.0c) spectro-polarimetric forward fitting \citep{Arnaud96}.  No significant polarization is detected in the full 2--8 keV IXPE band. With the PCUBE analysis, we have the normalized Stokes parameters $Q/I=-0.018 \pm 0.017$ and $U/I=0.015 \pm 0.017$ combining the three DUs, giving an upper limit of 5.3\% for MDP$_{99}$.

Spectro-polarimetric analysis is performed using \textsc{xspec} to jointly fit the three DUs in a two-step procedure. In the first step, the $I$ energy distribution is fitted with a spectral model. In the second step, the spectral model is fixed, while $U$ and $Q$ are fitted. This method thus does a joint forward folded fit of the Stokes fluxes to the binned $I$, $Q$, $U$ spectra with the fixed spectral model. We applied a constant energy binning of 200 eV for the $I$, $Q$, and $U$ data.

\begin{figure}[htbp]
\centering
\includegraphics[width=\linewidth]{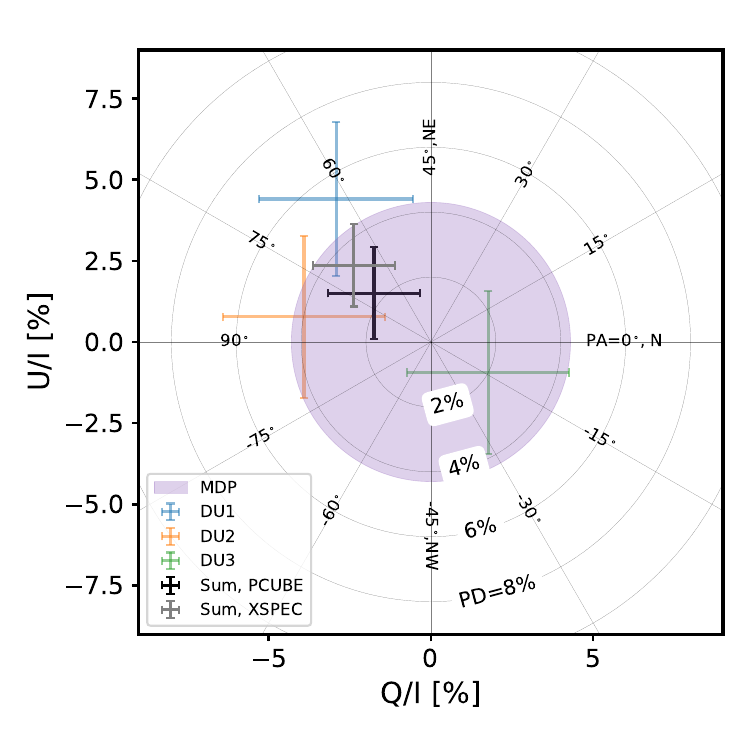}
\caption{Normalized Stokes parameters $Q/I$ and $U/I$ for different DU from phase-averaged analysis as measured with the \texttt{PCUBE} algorithm in \textsc{ixpeobssim} and \textsc{xspec}.}
\label{fig:over_q_u}
\end{figure}

The \psrb PWN binned $I$ spectra from the three DUs were fitted with the model \textsc{const*tbabs*powerlaw}, where \texttt{const} accounts for uncertainties in the absolute effective area of the three DUs, and \texttt{tbabs} takes into account the interstellar absorption. Here, we fixed the column density to $N_H=4.6\times10^{21}$\,cm$^{-2}$ as measured by \textit{Chandra} \citep{2001ApJ...546.1159K}.  The best-fit $I$ spectra and models for the three DUs are shown in the left panel of Fig.~\ref{fig:spectra_q_u}; best-fit values are reported in Table~\ref{tab:xspec_par}. These spectral fit values are in agreement with those of \cite{2001ApJ...546.1159K}. Fixing spectral parameters from $I$ and fitting with \texttt{polconst} of \textsc{xspec} provides the 2--8 keV $Q$ and $U$ spectra (Fig.~\ref{fig:spectra_q_u}, right panels). These 2--8 keV band-averaged, aperture-averaged (100$^{\prime \prime}$ radius) polarization degree and angle measurements are summarized in Fig.~\ref{fig:over_q_u}. The results are consistent between the three DUs and the two different analysis methods.

\begin{table}[ht]
\centering
\caption{\label{tab:xspec_par}Main results of the phase-averaged spectro-polarimetric analysis. Uncertainties are at 68\% CL.}
\begin{tabular}{crl}\hline \hline
    \multicolumn{3}{c}{\textsc{constant*TBabs*powerlaw}} \\
    \hline
    Model&Parameter&Value\\
    \hline
    \textsc{TBabs} & \textsc{N$_H$} [10$^{22}$ cm$^{-2}$]& 0.46 (frozen)  \\
    \hline
    \textsc{powerlaw} & $\Gamma$& 2.081$^{+0.014}_{-0.016}$  \\
    & \textsc{norm} [photon~keV$^{-1}$~cm$^{-2}$~s$^{-1}$]& 0.01594$^{+0.00024}_{-0.00027}$  \\
    \hline
    \multicolumn{3}{c}{Cross normalization factors} 
    \\& $C_{\rm DU1}$ & 1.0  (frozen)\\
    &$C_{\rm DU2}$ & 0.953$^{+0.002}_{-0.002}$  \\
    &$C_{\rm DU3}$ & 0.871$^{+0.002}_{-0.001}$  \\
    \hline
    \multicolumn{3}{c}{$\chi^2$/dof = 449.5/439 = 1.06} \\
\hline\hline

& &  \\

\hline\hline
\multicolumn{3}{c}{\textsc{polconst *constant*TBabs*powerlaw}} \\
\hline
\textsc{polconst} & \textsc{pd} [$\%$] & 3.3$^{+1.3}_{-1.3}$\\
\hline
& \textsc{pa} [$^\circ$] & 68$^{+11}_{-11}$\\
\hline
\multicolumn{3}{c}{$\chi^2$/dof = 248.6/259 = 0.96} \\
\hline
\hline
\end{tabular}
\end{table}

\subsection{Phase-resolved analysis}
\subsubsection{Off-pulse}
We can use phase-resolved analysis to decompose the PWN X-ray emission from that of the pulsar. For this analysis, the source and background regions are the same as for the phase-averaged treatment. We define the phase range $\phi$=0--0.35 as the off-pulse (Fig.~\ref{fig:qu_phase}). From prior IXPE Crab analysis, we have seen that the wide PSF wings place some photons in the background aperture. Therefore we take the nebula background from the same 0--0.35 phase window to minimize pulsar contamination. We do not detect a significant off-pulse (PWN) polarization in the full 2--8 keV IXPE range, with $Q/I=-0.049\pm0.025$, $U/I=0.020\pm0.025$, and a PD below the MDP$_{99}$ of 7.6\%. 

Then we performed an energy-dependent analysis by dividing the data into three energy ranges: 2--4 keV, 4--6 keV, and 6--8 keV. Results of the normalized Stokes parameters $Q/I$ and $U/I$ are listed in Table~\ref{tab:off-pulse-pol}. In the 4--6 keV range, the polarization degree $(24.5\pm 5.3)\%$ at PA = $(78.1\pm6.2)^{\circ}$ is detected at 4.6$\sigma$ significance. The angle is consistent with the $(79\pm5)^{\circ}$ optical PA \citep{1990ApJ...352..167C}.

\begin{table*}
    \centering
    \begin{tabular}{c|c|c|c||c}
    \hline \hline
         & 2--4 keV  & 4--6 keV   & 6--8 keV   & 2--8 keV \\
 \hline
 & \multicolumn{4}{c}{off-pulse}\\
    \hline
     Q/I & -0.032 $\pm$ 0.028   & -0.224 $\pm$ 0.053    &  0.191 $\pm$ 0.163    & -0.049 $\pm$ 0.025\\
     U/I &  0.015 $\pm$ 0.028   &  0.099 $\pm$ 0.053    & -0.245 $\pm$ 0.163    &  0.020 $\pm$ 0.025\\
\hline
 & \multicolumn{4}{c}{on-pulse}\\
     \hline
     Q/I &  0.018 $\pm$ 0.072   & 0.488 $\pm$ 0.131    &  -0.157 $\pm$ 0.329    & 0.112 $\pm$ 0.076\\
     U/I &  0.076 $\pm$ 0.072   &  0.108 $\pm$ 0.129    & 0.707 $\pm$ 0.338    &  0.148 $\pm$ 0.076\\
     \hline \hline
    \end{tabular}
    \caption{Normalized Stokes parameters of the measured polarization of the off-pulse window (100\asec aperture, $\phi$=0--0.35) and the on-pulse window (60\asec aperture, $\phi$=0.5--0.9) for different energy ranges.}
    \label{tab:off-pulse-pol}
\end{table*}

\subsubsection{On-pulse}
\begin{figure}
    \centering
    \includegraphics[width=\linewidth]{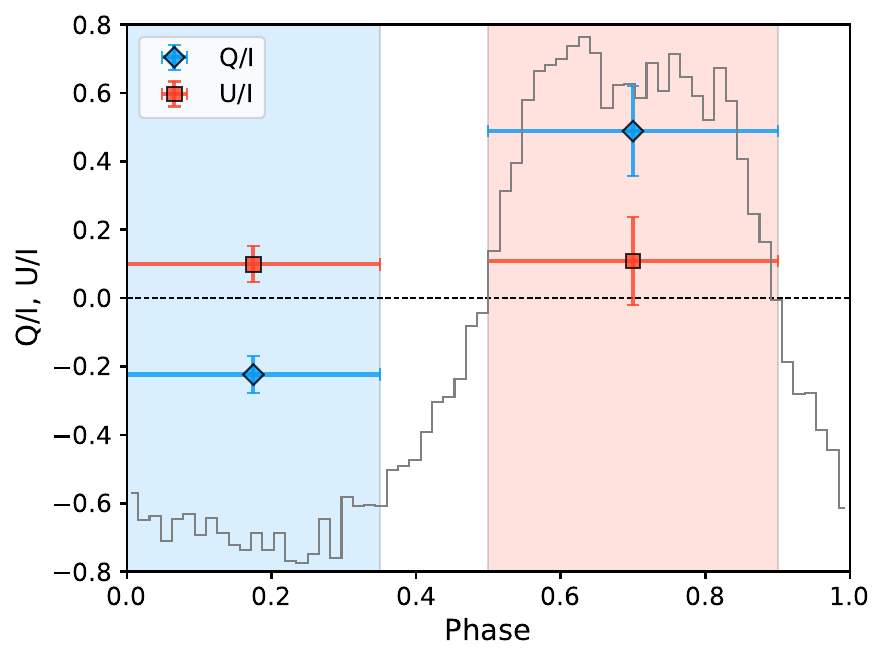}
    \caption{Polarization properties of the \psrb and its nebula in 4--6 keV energy band. Normalized Stokes parameters for the background-subtracted nebula emission (phase bin $\phi$=0--0.35 , in blue) and for the nebula-subtracted pulse emission ($\phi$=0.5--0.9, in red). The errors are for 1$\sigma$ standard deviation.}
    \label{fig:qu_phase}
\end{figure}

In a first analysis of the pulse phase polarization, we collected photons from phase range $\phi$=0.5--0.9 in a simple 60\asec aperture. For background, including nebula emission, we used photons from phase 0--0.35 (Fig~\ref{fig:qu_phase}). While no polarization is detected in the full 2--8 keV range, we do find a 3.8$\sigma$ detection in the 4--6 keV band, with results of the normalized Stokes parameters listed in Table~\ref{tab:off-pulse-pol}. In the 4--6 keV band, the $Q/I$ and $U/I$ before background subtraction are 0.070 $\pm$ 0.037 and 0.074 $\pm$ 0.037, and after background subtraction they are 0.488 $\pm$ 0.131 and 0.108 $\pm$ 0.129, resulting in a PD = $(50.0\pm13.1)\%$ and PA = $(6.2\pm7.4)^\circ$. Since the PWN is softer than the PSR and since residual particle background increasingly dominates at high energy, detection in an intermediate energy band is not unexpected.

Pulsar radio polarization often follows a rotating vector model (RVM) sweep, which could reduce the average PD in the broad on-pulse window. We attempted an RVM fit, but this did not significantly enhance the PD signal.

As a second analysis, we employed ``simultaneous" fitting as described in \cite{wong2023simulaneous}. This method uses external, {\it Chandra}-derived models for the spatial and spectral flux of the PWN and the phase-dependent PSR emission to assign weights to the PWN and PSR components in each of the several spatial and phase data bins. In this case, the 4--6 keV data were binned into ten 0.1-width phase windows and a 9$\times$9 grid of 10\asec pixels centered on the pulsar. The spatial binning helps separate the PSF-like PSR emission from the slightly broadened distribution from the PWN component. A simultaneous fit extracts the PWN and PSR contributions. 

To generate the energy-resolved nebula model, we passed an archival {\it Chandra} ACIS observation (ObsID 119) through \textsc{ixpeobssim v30.1} to account for the IXPE instrument response. Since we will add in a PSF-broadened phase-resolved pulsar component, we avoid double counting by removing the PSR point source from the image, excising a $r \sim 1.2$\asec region around the pulsar and replacing it with a sample of events from two regions, each $r \sim 0.9$\asec on either side of the excised region, using the average count rate. 

The \psrb light curve model was constructed from IXPE itself, by taking 2--8 keV photons within a 100\asec radius aperture, subtracting a background estimated from the 0.1--0.3 phase window, and binning into 50 equal-spaced phase bins. These counts were converted to specific flux ($\mathrm{s}^{-1}\ \mathrm{cm}^{-2}\ \mathrm{keV}^{-1}$) using the pulsed photon index measured by \cite{2001ApJ...546.1159K} $\Gamma = 1.83$, assumed to be constant with phase. This power-law flux model along with the ephemeris described above was passed through \textsc{ixpeobssim} to build the phase-resolved pulsar count model. A modest scaling was applied to both PSR and PWN components to match the total observed IXPE flux. We then made a least-squares fit to obtain the best-fit polarization parameters.

\begin{figure}
    \centering
    \includegraphics[width=\linewidth]{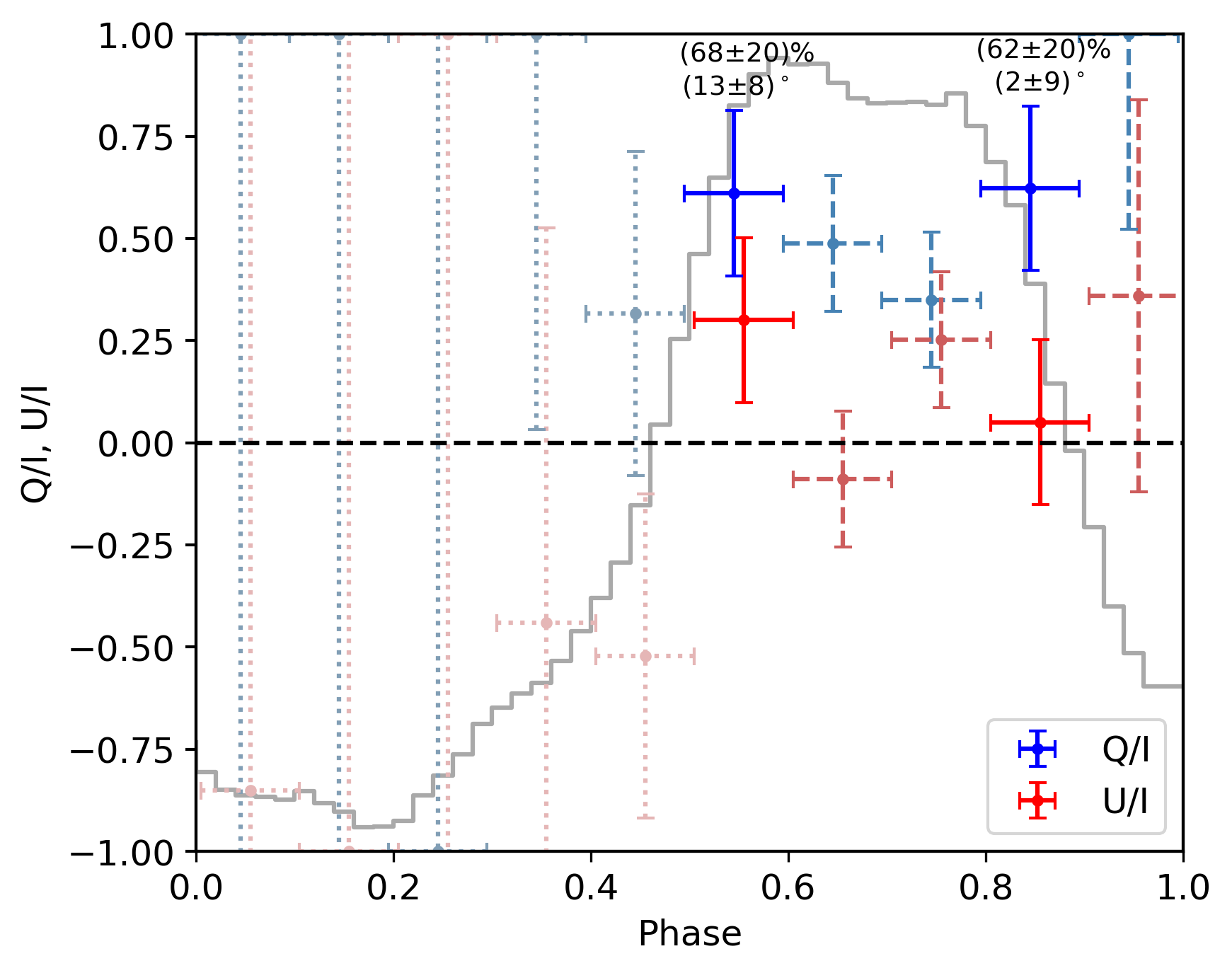}
    \caption{Phase-resolved polarization of \psrb in the 4--6 keV energy band. $>3\sigma$ significance bins are solid and $>2\sigma$ bins are dashed. Errors are 1$\sigma$ standard deviations. Gray light curve is displayed for reference. PD (\%) and PA ($^\circ$) are listed above the significant bins. Note that PD and PA for bins less than 3$\sigma$ have significant covariance, not included in the error bars.}
    \label{fig:simul_pulsar}
\end{figure}

\begin{figure}
    \centering
    \includegraphics[width=\linewidth]{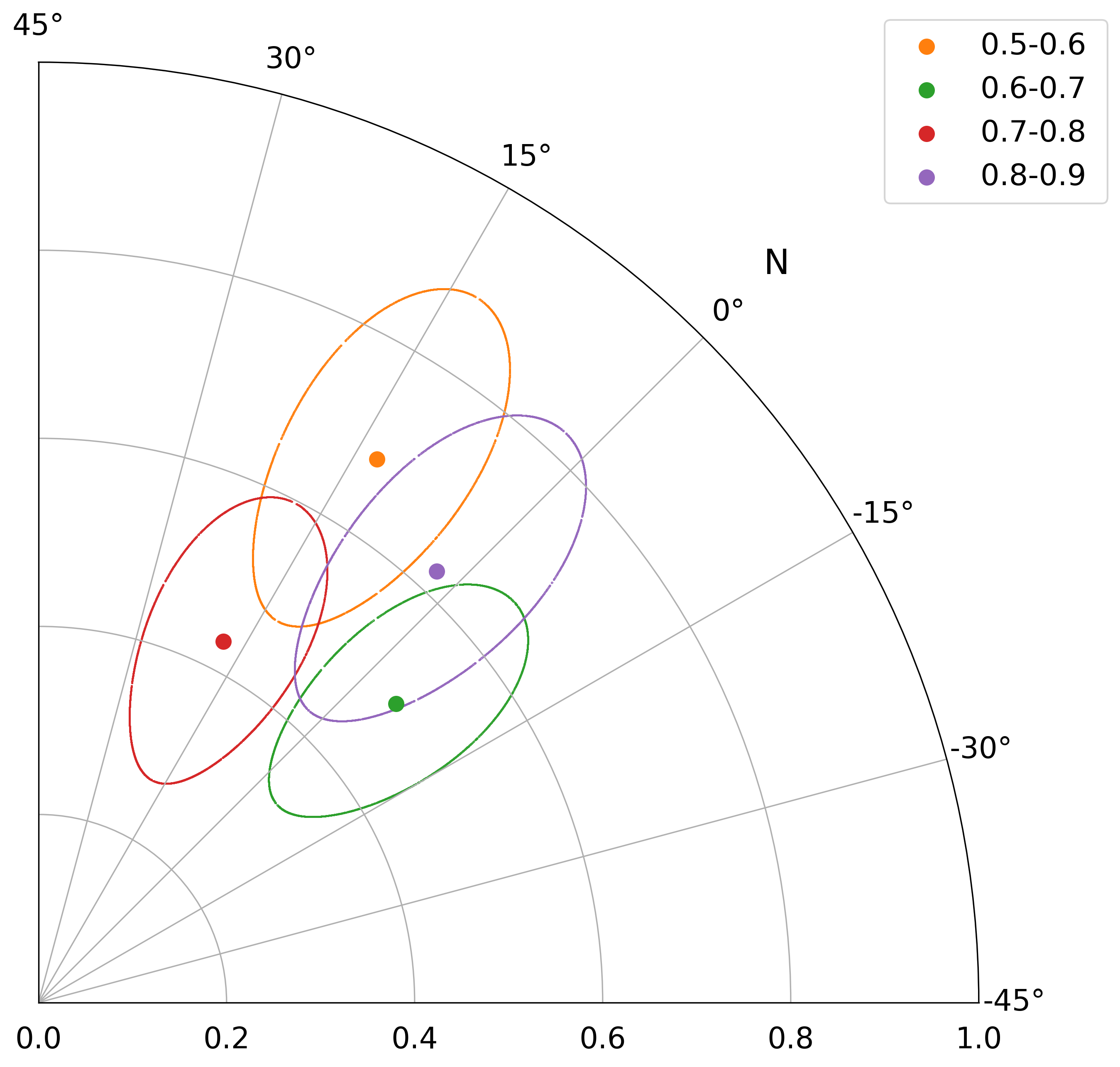}
    \caption{\psrb phase-resolved polarization measurements from the simultaneous fit for phase bins with PD$> 2.5\sigma$ detections. Contours show the 68\% CL.}
    \label{fig:simul_pulsar_contours}
\end{figure}

Figs.~\ref{fig:simul_pulsar} and \ref{fig:simul_pulsar_contours} and Table \ref{tab:pulsar_simul} display the pulsar polarization fit. We detect polarization at $\phi$=0.5--0.6 with PD=$(68.1\pm20.2)\%$ and $\phi$=0.8--0.9 with PD=$(62.4\pm20.1)\%$. These two phase bins are located at the boundaries of the broad peak, bracketing two 2$\sigma$ significance bins at lower ($\sim45\%$) PD. One of these is at 2.99$\sigma$ significance with PD=$(49.5\pm16.6)\%$. This polarization pattern has an interesting correspondence with the bifurcation of the peak, which \cite{dePlaa2003pulse} describe as a superposition of two Gaussian components with a separation of $\sim$0.2 for energies 2--20 keV. If these components are overlapping cones of emission with differing position angles, this could explain our PD results - mixing could cause the PD dip in the center of the broad peak. With significant measurements in only two phase bins, we cannot yet identify a definitive sweep. Additional exposure could measure the intervening bins, allowing useful model constraints. 

For the nebula, we did not find significant detection in any individual spatial pixel. However, we do find a significant polarization of the integrated nebula component, with PD=$(20.6 \pm 2.7)$\% and PA=$(82.8\pm 3.7)^{\circ}$. This is consistent with the values found using the 60\asec aperture and has a higher significance of $7.6\sigma$, in part because all phases can contribute to the nebula flux estimates.

\begin{table*}
    \centering
    \begin{tabular}{c|c|c|c|c|c|c|c|c}
    \hline \hline
     Phase  & Q/I  & Q/I err  & U/I  & U/I err  & PD & PD err & PA ($^\circ$) & Sig \\
    \hline
     0.5--0.6 & 0.61 & 0.20 & 0.30 & 0.20 & 0.68 & 0.20 & 13.1 & 3.37 \\
     0.6--0.7 & 0.49 & 0.17 & -0.09 & 0.17 & 0.50 & 0.16 & -5.13 & 2.99 \\
     0.7--0.8 & 0.35 & 0.17 & 0.25 & 0.17 & 0.43 & 0.16 & 17.9 & 2.60 \\
     0.8--0.9 & 0.62 & 0.20 & 0.05 & 0.20 & 0.62 & 0.20 & 2.29 & 3.10 \\
     \hline \hline
    \end{tabular}
    \caption{Normalized Stokes parameters for the pulsar phase-resolved polarization in the 4--6 keV energy band using ten equal-spaced phase bins obtained using simultaneous fitting, including only PD $>2.5\sigma$. Note that bins $< 3\sigma$ have significant PD-PA covariance. Refer to Fig.~\ref{fig:simul_pulsar_contours} for full 2D error contours.}
    \label{tab:pulsar_simul}
\end{table*}

\section{Discussion and conclusion}
\psrb is the fourth PWN observed by IXPE, after the Crab \citep{2023NatAs...7..602B}, Vela \citep{2022Natur.612..658X} and MSH 15-52 \citep{ixpeMSH1552}. All three PWNe are highly polarized, with local polarization degree reaching $>50$\%. \psrb's PWN is unresolved by IXPE, with a net phase-average polarization for the complex (PWN and PSR) below 5.3\% in 2--8 keV band. To separate the PWN and PSR components, we performed a phase-resolved analysis using two techniques. A simple on-off analysis allows us to get a statistically significant polarization detection from both the PWN and PSR in an optimized energy range 4--6 keV. 

For the PWN ($\phi$=0--0.35, Fig.~\ref{fig:qu_phase}), we measured PD=$(24.5\pm 5.3)\%$ and PA=$(78.1\pm6.2)^{\circ}$, detected at the 4.6$\sigma$ confidence level.
This PD is slightly higher than that of the Crab nebula. This may be due to \psrb's nearly edge-on PWN view (spin axis inclined $\zeta \sim 93^{\circ}$ to the line of sight versus $\zeta \sim$ 63$^{\circ}$ for the Crab). For such $\zeta$ a toroidal field projected to the sky has nearly constant position angle, reducing the de-polarization both due to line of sight integration, and spatial averaging over the PWN, with respect to the Crab. Indeed, local PD in the Crab PWN is found to be as high as 42\%, and locally higher values in the \psrb PWN are reported by \citet{2011MNRAS.413..611L} in spatially resolved optical polarimetry.
The measured PA of the PWN is consistent with the optical \citep{1990ApJ...352..167C} and radio \citep{2002ASPC..271..195D} values.
By fitting the {\it Chandra} X-ray PWN morphology, \citep{2008ApJ...673..411N} reported a \psrb spin axis position angle of 144.1$^{\circ}$. Surprisingly, the off-pulse PA is at large angle to this axis; one expects it to align well with the PWN symmetry axis, as for Crab and Vela. However, previous results on Crab and Vela, have shown that the most polarized regions are typically not close to the pulsar, where the magnetic field is generally oriented perpendicular to the PSR spin axis, but at the edge of the X-ray bright torus/rings, where environmental effects due to the interaction with the SNR ejecta, can lead to sizable deviations of the PA (about 20$^\circ$ in Crab).
Alternatively, the large deviation of $\sim$90$^{\circ}$ (78$^{\circ}$ vs. 144$^{\circ}$) in \psrb leads to a possibility that the brighter axis in the {\it Chandra} image selected as the torus in \cite{2008ApJ...673..411N} is the jet. \psrb could be a jet dominated system, like MSH 15-52, instead of a torus dominated one, like Crab or Vela.

For the pulsar ($\phi$=0.5--0.9, Fig.~\ref{fig:qu_phase}), we measured PD=$(50.0\pm13.1)\%$ and PA=$(6.2\pm7.4)^\circ$ at 3.8$\sigma$ confidence level, treating the broad pulse emission as a single phase bin. This result is improved by using the ``simultaneous'' fitting method \citep{wong2023simulaneous}, with which we detected polarization in the two 0.1-width phase bins bracketing the pulse, as shown in Fig.~\ref{fig:simul_pulsar}. These bins had PD=$(68\pm20)\%$ at $\phi$=0.5--0.6 and PD=$(62\pm20)\%$ at $\phi$=0.8--0.9. The intervening bins are at lower PD and 2--3$\sigma$ significance.
This has an interesting correlation with the analysis done by \cite{dePlaa2003pulse}, which purports that the pulse emission is the sum of two Gaussian components. More phase bins are needed to fully resolve the polarization structure in the pulse, but these two observations already hint at two distinct radiation components, perhaps at separate sites in the magnetosphere with different polarization, combining to make the main pulse emission. Within striped wind emission models \citep{petri}, a single large pulse, can be achieved only if the inclination of the spin axis with respect to the line of sight is close to the magnetic axis inclination. In this case, the core of the pulse is expected to be unpolarized, with polarization present only at the leading and trailing edges.
The ``simultaneous" fitting technique also recovers a consistent measurement of the nebula polarization at higher significance.

\psrb is similar to Crab in many respects, but the pulse profile is very different. Here, we see a broad pulse while the Crab profile shows two sharp peaks. \cite{2023NatAs...7..602B} reported a pulsed Crab PD detection in a very narrow phase range ($\Delta \phi$ = 0.02) at the main peak, while \citet{wong2023simulaneous} found significant polarization in several near-peak bins. Many models have been proposed for the high energy pulsar emission, but the radiation site is still not fully understood. Polarization could be a powerful tool to study this radiation by measuring the change of polarization angle with phase. At present, with polarization in only a few phase bins for \psrb, the Crab pulsar \citep{wong2023simulaneous} and PSR B1509-58 \citep{ixpeMSH1552}, it is difficult to test the models. Additional IXPE observations of these interesting sources could enable model discrimination.

\section*{Acknowledgments}
The Imaging X-ray Polarimetry Explorer (IXPE) is a joint US and Italian mission. The US contribution is supported by the National Aeronautics and Space Administration (NASA) and led and managed by its Marshall Space Flight Center (MSFC), with industry partner Ball Aerospace (contract NNM15AA18C).  The Italian contribution is supported by the Italian Space Agency (Agenzia Spaziale Italiana, ASI) through contract ASI-OHBI-2022-13-I.0, agreements ASI-INAF-2022-19-HH.0 and ASI-INFN-2017.13-H0, and its Space Science Data Center (SSDC) with agreements ASI-INAF-2022-14-HH.0 and ASI-INFN 2021-43-HH.0, and by the Istituto Nazionale di Astrofisica (INAF) and the Istituto Nazionale di Fisica Nucleare (INFN) in Italy.  This research used data products provided by the IXPE Team (MSFC, SSDC, INAF, and INFN) and distributed with additional software tools by the High-Energy Astrophysics Science Archive Research Center (HEASARC), at NASA Goddard Space Flight Center (GSFC). 

This work is supported by National Key R\&D Program of China (grant No. 2023YFE0117200) and National Natural Science Foundation of China (grant No. 12373041 and grant No. 12133003). 
C.Y.N. and Y.J.Y. are supported by a GRF grant of the Hong Kong Government under HKU 17305419. 
Funding for this work was provided in part by contract NNM17AA26C from the MSFC to Stanford and 80MSFC17C0012 to MIT in support of the {\it IXPE} project.
N.B. is supported by the INAF MiniGrant “PWNnumpol - Numerical Studies of Pulsar Wind Nebulae in The Light of IXPE”.
I.L. is supported by the NASA Postdoctoral Program at the Marshall Space Flight Center, administered by Oak Ridge Associated Universities under contract with NASA.
W.C.G.H. acknowledges support through grant 80NSSC23K0078 from NASA.
This research has made use of \nicer data. We thank \nicer staff for the scheduling of these observations. 
This paper employs a list of Chandra datasets, obtained by the Chandra X-ray Observatory, contained in \dataset[DOI: https://doi.org/10.25574/cdc.177]{https://doi.org/10.25574/cdc.177}.

\bibliographystyle{aasjournal}

\end{document}